\documentclass[graybox]{svmult}

\usepackage{mathptmx}       
\usepackage{helvet}         
\usepackage{courier}        
\usepackage{makeidx}         
\usepackage{graphicx}        
\usepackage{multicol}        
\usepackage[bottom]{footmisc}

\usepackage{graphicx,color}
\usepackage{amsfonts,amsmath,amssymb}

\makeindex             

\newcommand{\1}{{\rm 1}\kern-0.24em{\rm I}}
\newcommand{\R}{{\rm I}\kern-0.20em{\rm R}}
\renewcommand{\P}{{\rm I}\kern-0.20em{\rm P}}
\renewcommand{\E}{\mbox{E}}

\newcommand{\BEQS}{\begin{equation*}}
\newcommand{\EEQS}{\end{equation*}}
\newcommand{\BEAS}{\begin{eqnarray*}}
\newcommand{\EEAS}{\end{eqnarray*}}
\newcommand{\BEQ}{\begin{equation}}
\newcommand{\EEQ}{\end{equation}}
\newcommand{\BEL}[1]{\begin{equation}\label{#1}}
\newcommand{\BEA}{\begin{eqnarray}}
\newcommand{\EEA}{\end{eqnarray}}
\newcommand{\bi}{\begin{itemize}}
\newcommand{\ei}{\end{itemize}}

\newcommand{\auccd}{\mbox{AUC}^{\mathbb{C},\mathbb{D}}}
\newcommand{\aucis}{\mbox{AUC}^{\mathbb{I},\mathbb{S}}}

\newcommand{\dynamic}{\emph{dynamic}}
\newcommand{\cumulative}{\emph{cumulative}}

\begin{document}

\title*{Time-dependent AUC with right-censored data: a survey study}
\author{Paul Blanche, Aur\'elien Latouche, Vivian Viallon}
\institute{Paul Blanche \at Univ. Bordeaux, ISPED and INSERM U897, Bordeaux, FRANCE \email{Paul.Blanche@isped.u-bordeaux2.fr}
\and Aur\'elien Latouche \at Conservatoire national des arts et m\'etiers, Paris, FRANCE \email{aurelien.latouche@cnam.fr}
\and Vivian Viallon \at UMRESTTE (Univ. Lyon 1 and IFSTTAR), Bron, FRANCE \email{viallon@math.univ-lyon1.fr}}
%
%
\maketitle

\vspace{-2cm}
\abstract{The ROC curve and the corresponding AUC are popular tools for the evaluation of diagnostic tests. They have been recently extended
to assess prognostic markers and predictive models. However, due to the many particularities of time-to-event outcomes, various definitions and estimators
have been proposed in the literature. This review article aims at presenting the ones that accommodate to right-censoring, which is common when evaluating
such prognostic markers.}

\section{Introduction}\label{sec:Intro}

In the medical literature, a variety of general criteria have been used to assess diagnostic tests \cite{pepe2004statistical, gail2005criteria}. Among them, the ROC curve and the area under it -- the AUC -- are popular tools, originally aimed at evaluating the discriminant power of continuous diagnostic tests. In this simple situation, the outcome status $D$ is a binary variable (typically, $D=1$ for cases and $D=0$ for controls) and the ROC curve for a continuous diagnostic test $X$ plots the true positive rate, or sensitivity, ${\rm TPR}(c)=\P(X>c|D=1)$  against the false positive rate, or one minus the specificity, ${\rm FPR}(c)=\P(X>c|D=0)$, when making threshold $c$ vary. The AUC, which is the area under this curve, is a commonly used summary measure of the information contained in the sequences $({\rm TPR})_{c\in\R}$ and $({\rm FPR})_{c\in\R}$. As such, it inherits some of the properties of true and false positive rates. In particular, it is not affected by the disease prevalence (unlike positive and negative predictive values) and it can be evaluated from random samples of cases and controls. It also has a nice interpretation since it corresponds to the probability that the marker value of a randomly selected case exceeds that of a randomly selected control.

\noindent The extension of the AUC (and other evaluation criteria) to the setting of prognostic markers has raised several issues. In particular, when evaluating such markers, the outcome status typically changes over time: in a cohort study for instance, patients are diseased-free when entering the study and may develop the disease during the study. 
This leads to three major differences with the evaluation of diagnostic tests.
 First, this time-dependent outcome status (which may be defined in several ways, as will be seen in Section \ref{sec:VV}) naturally implies that sensitivity, specificity, ROC curves, their AUC values and, more generally, any extension of the criteria used in the diagnostic setting, are functions of time as well. Second, the time-to-event, i.e. the time between the entry in the study and the disease onset, is usually \emph{censored} and not fully observed, requiring dedicated inference.
 Third, the time lag between the entry in the study and the disease onset also leads to two further refinements: $(i)$ the marker can be repeatedly measured over time and $(ii)$ competing events (in addition to censoring) may be observed between the entry and the putative disease onset.

\noindent These particularities has  led to the development of numerous methods aimed at estimating the time-dependent AUC for prognostic markers. In this paper, we review those that accommodate to right-censoring. Some notations are introduced in the following Section \ref{sec:Notations}. Then, in Section \ref{sec:VV} we will present several definitions and estimators of the time-dependent AUC in the ``standard'' setting of a baseline marker and univariate survival data. Section \ref{sec:PB} will cover the case of longitudinal markers which corresponds to the marker being repeatedly measured over time, while we will discuss the competing events setting in Section \ref{sec:AL}. Finally, concluding remarks will be given in Section \ref{sec:Discussion}.
\section{Notations}\label{sec:Notations}
Let $T_i$ and $C_i$  denote survival and censoring times
for subject $i$, $i=1,\ldots,n$. We further let
$Z_i=\min(T_i, C_i)$ and $\delta_i=\1(T_i\leq C_i)$ denote the observed time  and the status indicator respectively. We will 
denote by $D_i(t)$ the time-dependent outcome status for subject
$i$ at time $t$, $t\geq0$. Several definitions for $D_i(t)$ will be given
hereafter, but we will always have $D_i(t)=1$ if subject $i$ is considered as
a case at time $t$ and $D_i(t)=0$ if subject $i$ is considered as a control at time $t$.

\noindent We will denote by $X$ the marker under study, which can be a single biological marker or several biological markers combined into a predictive model (in this case, it is assumed throughout this article that the predictive model has been constructed on an independent data set; otherwise sub-sampling techniques are needed \cite{lee12}). Without loss of generality, we will suppose that larger values of $X$ are associated 
with greater  risks (otherwise, $X$ can be recoded to achieve this). 
We will denote by $g$ and $G^{-1}$ the probability density function and the quantile function of marker $X$.  In Section \ref{sec:VV}, we assume that marker $X$ is measured once at $t=0$, and we will denote 
by $X_i$ the marker value for subject $i$. In Section \ref{sec:PB}, which treats the longitudinal setting, the marker is measured repeatedly over time, and we will denote 
by $X_i(s)$ the marker value at time $s$ for subject $i$. 


\section{Time-dependent AUCs in the standard setting}\label{sec:VV}

Definitions of time-dependent ROC curves, ROC$(t)$, follow from definitions of usual ROC curves and thus rely on first defining time-dependent true and false positive rates.  For any threshold $c$, these two functions of time are defined as $\mbox{TPR}(c,t) =
\P(X>c|D(t)=1)$ and $\mbox{FPR}(c,t) = \P(X>c|D(t)=0)$. ROC$(t)$ then simply plots $\mbox{TPR}(c,t)$ against $\mbox{FPR}(c,t)$ making threshold $c$ vary. The time-dependent AUC at time $t$ is then defined as the area under this curve, 
\begin{equation}\label{Eq_AUC}
\mbox{AUC}(t)= \int_{-\infty}^\infty
\mbox{TPR}(c,t)\left|\frac{\partial \mbox{FPR}(c,t)}{\partial c}\right|dc.
\end{equation}

\noindent As a matter of fact, these definitions deeply rely on that of the outcome status at time $t$, $D(t)$. Heagerty and Zheng \cite{heagerty05} described 
several definitions of {\em cases} and {\em controls}  in this survival outcome setting. According to Heagerty and Zheng's terminology and
still denoting by $T_i$ survival time for subject $i$, cases are said to be \emph{incident} if $T_i=t$ is used to define cases at time $t$, and \emph{cumulative} if $T_i\leq t$ is used instead. Similarly, depending on whether $T_i>\tau$ for a large time $\tau> t$ or $T_i>t$ is used for defining controls at time $t$, they are said to be \emph{static} or \emph{dynamic} controls. Depending on the definition retained for cases and controls at time $t$, four definitions of the time-dependent AUC value may be put forward. In the following paragraphs, we will present formulas and estimators for the most commonly used ones and will discuss their respective interests.

\subsection{The cumulative dynamic AUC: $\auccd(t)$}

The setting of cumulative cases and dynamic controls may be regarded as the most natural choice for planning enrollment criteria in clinical trials or when specific evaluation times are of particular interest. It simply corresponds to defining cases at time $t$ as subjects who experienced the event prior to time $t$, and controls at time $t$ as patients who were still event-free at time $t$. In other words, it corresponds to setting $D_i(t)= \1(T_i\leq t)$. Cumulative
true positive rates and dynamic false positive rates are then respectively defined as
\begin{equation}
\mbox{TPR}^{\mathbb C}(c,t)=\P(X>c|T\leq t)\quad{\rm and}\quad
\mbox{FPR}^{\mathbb D}(c,t)=\P(X>c|T> t).
\label{cumTPRdynFPR}
\end{equation}
The \cumulative/\dynamic\ AUC at time $t$ is then obtained by using these definitions of true and false positive rates  in (\ref{Eq_AUC}). Usually, however, $\1(T\leq t)$ is not observed for all subjects due to the presence of censoring before time $t$ and simple contingency tables can therefore not be used to return estimates of $\mbox{TPR}^{\mathbb C}(c,t)$, $\mbox{FPR}^{\mathbb D}(c,t)$ and $\auccd(t)$. To handle censoring, Bayes' theorem can be used to rewrite $\auccd(t)$ as a function of the conditional survival function $\P(T> t|X=x)$ (see paragraph \ref{modelbased} below). Other approaches rely on so-called Inverse Probability of Censoring Weighted (IPCW) estimates (see paragraph \ref{IPCW} below). Before describing these two approaches in more details below, we shall add that Chambless and Diao \cite{chambless06} developed an alternative method -- which will not be described here -- based on an idea similar to the one used to derive the Kaplan-Meier estimator of the cumulative distribution function in the presence of censoring. Among all these methods, only those relying on primary estimates of $\P(T> t|X=x)$ (and a recent extension of  IPCW estimates proposed in \cite{Paul}) may account for the dependence between censoring and the marker (since they basically only assume that $T$ and $C$ are independent given $X$ and not that $T$ and $C$ are independent). We refer the reader to \cite{Paul,CJS:CJS10046,viallon11} for empirical comparisons  and illustrations of these various methods.

\subsubsection{Methods based on primary estimates of $\P(T> t|X=x)$}\label{modelbased}

Bayes' theorem yields the following expressions for  $\mbox{TPR}^{\mathbb C}(c,t)$ and $\mbox{FPR}^{\mathbb D}(c,t)$
\[ \mbox{TPR}^{\mathbb C}(c,t) = \frac{\int_c^\infty\P(T\leq t|X=x)g(x)dx}{\P(T\leq t)},\quad \!\!\mbox{FPR}^{\mathbb D}(c,t) =  \frac{\int_c^\infty\P(T> t|X=x)g(x)dx}{\P(T> t)}\cdot\]
From (\ref{Eq_AUC}),  it readily follows that 
\begin{equation}\label{AUCCD_Bayes}
\auccd(t)=\int_{-\infty}^\infty\int_c^\infty
\frac{\P(T\leq t|X=x)\P(T> t|X=c)}{\P(T\leq t)\P(T>t)}g(x)g(c)dxdc.
\end{equation}
Since $\P(T> t) = \int_{-\infty}^\infty \P(T> t|X=x)g(x)dx$, any estimator $\widehat{S_n}(t|x)$ of the conditional survival function $\P(T> t|X=x)$ yields an estimator of $\auccd(t)$:
\begin{equation*}
\widehat{\mbox{AUC}}^{\mathbb{C},\mathbb{D}}(t) = \frac{\sum_{i=1}^n\sum_{j=1}^n \widehat{S_n}(t|X_j)[1-\widehat{S_n}(t|X_i)]\1(X_i>X_j)}{\sum_{i=1}^n\sum_{j=1}^n \widehat{S_n}(t|X_j)[1-\widehat{S_n}(t|X_i)]}.
\end{equation*}
In \cite{chambless06}, the authors suggested to use a Cox model to derive estimates  $\widehat{S_n}(t|x)$, while one of the methods described in Heagerty \emph{et al.} \cite{heagerty00} reduces to using the conditional Kaplan-Meier estimator as in \cite{Akritas94}. Some theoretical results for these two methods can be found in \cite{Song08} and \cite{cai11,Chiang10,Hung11}  respectively. 

\noindent We shall add that Viallon and Latouche \cite{viallon11} related $\auccd(t)$ to the quantity $\P(T\leq t|X=G^{-1}(q))$ -- a time-dependent version of the predictiveness curve:
\begin{equation*}
\auccd(t)=\frac{\int_{0}^{1} q
\P(T\leq t|X=G^{-1}(q))dq-[\P(T\leq t)]^2/2}{\P(T>t)\P(T\leq t)}\cdot
\label{Eq_AUC_CD_Predictiveness}
\end{equation*}
This confirms that most standard statistical summaries of predictability and discrimination can be derived from the predictiveness curve, as pointed out in \cite{gail2005criteria,GuPepe2009}.

\subsubsection{Methods based on IPCW estimators}\label{IPCW}

In  \cite{CJS:CJS10046} and \cite{uno07}, the authors independently suggested to use IPCW-type estimates: 
\begin{equation*}
\widehat{\mbox{TPR}}^{\mathbb C}(c,t) = \frac{\sum_{i=1}^n \1(X_i >c, Z_i\leq t)\frac{\delta_i}{n\hat S_C(Z_i)}}{\sum_{i=1}^n \1(Z_i\leq t)\frac{\delta_i}{n\hat S_C(Z_i)}}, \quad
\widehat{\mbox{FPR}}^{\mathbb D}(c,t)= \frac{\sum_{i=1}^n \1(X_i >c, Z_i>t)}{\sum_{i=1}^n \1(Z_i> t)},\label{Uno_TPR_FPR}
\end{equation*}
where $\hat S_C(\cdot)$ is the Kaplan-Meier estimator of the survival function of the censoring time $C$. The expression of the false positive rate estimator is more compact because weights all equal $1/(n\hat S_C(t))$ under the assumption of independence between $C$ and $X$, and then vanish. $\widehat{\mbox{FPR}}^{\mathbb D}(c,t)$ corresponds to $1$ minus the empirical distribution function of $X$ among individuals for whom $Z_i>t$. In the absence of censoring before time $t$, $\widehat{\mbox{TPR}}^{\mathbb C}(c,t)$ also reduces to the usual empirical version of ${\mbox{TPR}}^{\mathbb C}(c,t)$, \emph{i.e.}, $1$ minus the empirical distribution function of $X$ among individuals for whom $T_i\leq t$. 

\noindent It can be shown (see \cite{CJS:CJS10046,satten2001kaplan}) that an estimator of $\auccd(t)$ is then given by
\begin{equation*}
\widehat{\mbox{AUC}}^{\mathbb{C},\mathbb{D}}(t)= \frac{\sum_{i=1}^n\sum_{j=1}^n \1(Z_i\leq t)\1(Z_j>t)\1(X_i>X_j)\frac{\delta_i}{\hat S_C(Z_i)\hat S_C(t)}}{n^2 \hat  S(t) [1-\hat S(t)]},
\end{equation*}
where $ \hat  S(t)$ is the Kaplan-Meier estimator of $\P(T>t)$. \vskip5pt

\noindent Theoretical guarantees for these estimators can be found in \cite{CJS:CJS10046} and \cite{uno07}. These estimators are in a sense more flexible than those presented in paragraph \ref{modelbased} above: they are model-free and they do not rely on any bandwidth selection (unlike the estimator of Heagerty \emph{et al.} \cite{heagerty00} for instance, which is based on a local version of the Kaplan-Meier estimator). However, when censoring may depend on marker $X$, quantities like the conditional survival function of $C$ given the marker $X$, $S_C(\cdot | X)$, have to be estimated   \cite{Paul}, which implies either to work under some (semi-)parametric model or the selection of some parameter if nonparametric estimation is prefered.

\subsection{The incident static AUC: $\aucis(t)$}

Using the dynamic definition of controls, the control group varies with time, and so does the $x$-axis of the corresponding
ROC curves: in situations where trends over time are of particular interest, this renders their interpretation more difficult (since such trends may be partly due to changing control groups).
Moreover, the group of static controls is interesting in that it tries to mimic the group of individuals who never develop the disease, which can be seen as an ideal control group in some situations. In particular, patients with preclinical diseases are eliminated from the control group as far as possible, if $\tau$ is large enough.  

\noindent Regarding cases, the incident definition has several advantages over the dynamic definition \cite{Pepe08}. The cumulative TPR does not distinguish between events that occur early versus late, and it shows redundant information over time (since early events are also included in the cumulative TPR for late evaluation times). Moreover, as pointed out by \cite{cai06}, the cumulative TPR can be computed from the incident TPR when the distribution of the event time is known.

\noindent Putting all this together, several authors have proposed estimators of the time-dependent ROC curve relying on the incident definition of cases and static definition of controls. Standard numerical integration techniques are then used to compute an estimate for $\aucis(t)$ from the estimators of the ROC curve. \vskip5pt

\noindent Incident true positive rates and static false positive rates are defined as 
\begin{equation}
\mbox{TPR}^{\mathbb I}(c,t)=\P(X>c|T = t)\quad{\rm and}\quad
\mbox{FPR}^{\mathbb S}_\tau(c)=\P(X>c|T>\tau).
\label{incTPRstaFPR}
\end{equation}
Applying Bayes' theorem, they can further be rewritten (see, \emph{e.g.}, \cite{Song08})
\begin{equation*}
\mbox{TPR}^{\mathbb I}(c,t)=\frac{\int_c^\infty f(t|x)g(x) dx}{\int_{-\infty}^\infty f(t|x)g(x) dx}\quad{\rm and}\quad
\mbox{FPR}^{\mathbb S}_\tau(c)= \frac{\int_c^\infty \P(T> \tau| X=x )g(x) dx }{\int_{-\infty}^\infty \P(T> \tau| X=x )g(x) dx},
\end{equation*}
where $f(t|x) = \partial \P(T\leq t| X=x )/\partial t$ is the conditional density function of $T$ given $X=x$. 
\noindent Under a standard Cox model of the form $\lambda(t;X) = \lambda_0(t)\exp(\beta X)$ -- here $\lambda(t;X)$ stands for the conditional hazard rate of $T$ given $X$ while $\lambda_0$ is the unspecified baseline hazard rate -- Song and Zhou  \cite{Song08} deduced that 
\begin{eqnarray*}
\mbox{TPR}^{\mathbb I}(c,t)&=&\frac{\int_c^\infty \exp(\beta x)\exp\{-\Lambda_0(t) \exp(\beta x)\}g(x) dx}{\int_{-\infty}^\infty \exp(\beta x)\exp\{-\Lambda_0(t) \exp(\beta x)\}g(x) dx}\\
\mbox{FPR}^{\mathbb S}_\tau(c)&=& \frac{\int_c^\infty \exp\{-\Lambda_0(\tau) \exp(\beta x)\}g(x) dx }{\int_{-\infty}^\infty \exp\{-\Lambda_0(\tau) \exp(\beta x)\}g(x) dx},
\end{eqnarray*}
where $\Lambda_0(t) = \int_{-\infty}^t\lambda_0(u)du$ is the cumulative baseline hazard function.
 Estimation of $\mbox{TPR}^{\mathbb I}(c,t)$ and $\mbox{FPR}^{\mathbb S}_\tau(c)$ can then be achieved by plug-in methods. We shall add that Song and Zhou actually considered a slightly more general set-up where additional covariates can be accounted for. \vskip5pt

\noindent In \cite{heagerty05}, Heagerty and Zheng adopted a slightly different approach. To estimate $\mbox{TPR}^{\mathbb I}(c,t)$, they used a (possibly time-varying-coefficients) Cox model of the form $\lambda(t;X) = \lambda_0(t)\exp(\beta(t) X)$ in combination with the fact that the distribution of $X\cdot$ $\exp(\beta X)$ for subjects in the risk set at time $t$ is equal to the conditional distribution of $X$ given $T=t$ (see, \emph{e.g., } \cite{XuOquigley}). Setting $R(t)=\{i: Z_i\geq t\}$, this leads to 
\[ \widehat{\mbox{TPR}}^{\mathbb I}(c,t) = \frac{\sum_{i \in R(t)} \1(X_i>c)\exp\{\beta(t)X_i\}}{\sum_{i\in R(t)}\exp\{\beta(t)X_i\}}.\]
\noindent As for the estimation of  $\mbox{FPR}^{\mathbb S}_\tau(c)$, they proposed a model-free approach using the empirical distribution function for marker values among the control set $S_\tau:=\{i: Z_i>\tau \}$. Namely, denoting by $n_\tau$ the cardinality of $S_\tau$, they proposed
\[\widehat{\mbox{FPR}}^{\mathbb S}_\tau(c) = \frac{1}{n_\tau} \sum_{i\in S_\tau} \1(X_i> c),\]
which is $\widehat{\mbox{FPR}}^{\mathbb D}(c,t)$ of Section \ref{IPCW}, except $\tau$ is used instead of $t$.

\noindent  Cai \emph{et al.} \cite{cai06} proposed another approach in the context of longitudinal markers; it will be described in more details in Section \ref{sec:PB}. In addition, two non parametric approaches were recently proposed (see \cite{SZ12} and \cite{saha2012non}).

\noindent Note also that estimators for the time-dependant incident/dynamic  AUC, AUC$^{{\mathbb I},{\mathbb D}}(t)$, can be obtained by simply replacing $\tau$ by $t$ in the definitions of  $\widehat{\mbox{FPR}}^{\mathbb S}_\tau(c)$ above \cite{heagerty05}. A global accuracy measure has further been derived from the definition of AUC$^{{\mathbb I},{\mathbb D}}(t)$, which is particularly appealing when no a priori time $t$ is identified and/or when trends over time are not of interest  \cite{heagerty05}.

\section{Time-dependent ROC curve and AUCs with longitudinal marker}\label{sec:PB}


 In this section, we review extensions of the above estimators for longitudinally collected subject measurements. For instance some authors would assess the discrimination performance of CD4 counts repeatedly measured every week on time from seroconvertion to progression to AIDS \cite{zheng07}. Therefore, time-dependent sensitivities and specificities have been extended to deal with the fact that $(i)$ the time at which marker $X$ is measured can vary and $(ii)$ marker can be repeatedly  measured on the same subject.  Let $s$ denote the timing of marker measurement and $X(s)$ the marker value at time $s$. For $t \geq s$, Zheng and Heagerty \cite{zheng07} extended \emph{cumulative}/\emph{dynamic} definitions
\begin{align*}
&\mbox{TPR}^{\mathbb C}(c,s,t)=\P(X(s)>c|T \in [s,t] ), \qquad \mbox{FPR}^{\mathbb D}(c,s,t)= \P(X(s)>c|T>t ).
\end{align*} 
For a fixed time $\tau \geq s$, Zheng and Heagerty \cite{zheng2004} extended  \emph{incident}/\emph{static} definitions:
\begin{align*} 
& \mbox{TPR}^{\mathbb I}(c,s,t)=\P(X(s)>c|T=t) \quad \mbox{and} \quad \mbox{FPR}^{\mathbb S}(c,s,\tau)=\P(X(s)>c|T> \tau).
\end{align*}



\noindent Although others approaches have been proposed to estimate these quantities, we only review estimators that deal with censored data here. We should also mention that in this longitudinal context, estimators of the AUC are obtained by numerically integrating estimators of the ROC curve.

\subsection{Cumulative dynamic estimators with longitudinal marker}

Rizopoulos \cite{rizopoulos11} recently proposed a joint modeling approach. The marker trajectory is modeled by usual linear mixed model for longitudinal data, and a parametric proportional hazard is used to model the time-to-event given the marker trajectory. 
The two submodels are linked with  shared random effects to capture the intra-subject correlation. Standard maximum likelihood estimation is  used to fit the joint model.  Then, $\mbox{TPR}^{\mathbb C}$ and $\mbox{FPR}^{\mathbb D}$ are computed from the estimated parameters and Monte Carlo simulations are used to make inference. As this approach is fully parametric, its main advantage is its efficiency. This approach also allows censoring to depend on the marker \cite{tsiatis2004}. The counterpart is that the parametric model must be carefully chosen, and checking model fit is not straightforward.

\noindent  A more flexible methodology was proposed in \cite{zheng07}, with fewer parametric assumptions. Setting $T^*=T-s$, the ``residual failure time", and $t^{*}=t-s$, they rewrote
\begin{align*}
\mbox{TPR}^{\mathbb C}(c,s,t^*)=\frac{1-F_{X\vert s}(c) - S(c,t^* \vert s)}{1-S(t^* \vert s)}, \qquad
\mbox{FPR}^{\mathbb D}(c,s,t^*)= 1- \frac{S(c,t^* \vert s)}{S(t^* \vert s)},
\end{align*}
with $F_{X \vert s}(c)=\P(X(s)<c|s,T^*>0 )$ the conditional distribution of marker given measurement time, $S(t^* \vert s)=\P(T^*>t^* \vert s, T^*>0 )$ the survival probability for individuals who survived beyond $s$ and $S(c,t^* \vert s)=\P(X(s)>c,T^*>t^* \vert s, T^*>0 )$. They proposed to estimate $F_{X \vert s}(c)$ with the semiparametric estimator proposed by Heagerty and Pepe \cite{heagerty1999}. Therefore, only the location and scale of the conditional distribution of marker given measurement time  are parametrized. To estimate the survival terms $S(c,t^* \vert s)$ and $S(t^* \vert s)$, they proposed the use of a  ``partly conditional" hazard function to model the residual failure time $T^*=T-s$. For  subject $i$ at measurement time $s_{ik}$, this  function is modeled by
\begin{displaymath}
\lambda_{ik}(t^* \vert X_i(s_{ik}),0\leq s_{ik} \leq T_i)=\lambda_0(t^{*})\exp\left[ \beta(t^*)X_i(s_{ik}) + \alpha^T f(s_{ik}) \right]
\end{displaymath}
where $f(s)$ are vectors of spline basis functions evaluated at measurement time $s$, and $\lambda_0(t^{*})$ is left unspecified. Estimators of $\beta(\cdot)$ and $\alpha$ have been previously proposed \cite{zheng2005}. As this approach is semiparametric, its main advantage is its flexibility. However, by contrast to the approach of  Rizopoulos \cite{rizopoulos11}, this one is less efficient and does not allow marker-dependent censoring.

\subsection{Incident static estimators with longitudinal  marker}
Several authors consider the incident/static definition of AUC \cite{slate2000,etzioni1999,cai06,subtil09}. However, censored data are only accounted for by   Cai \emph{et al.} \cite{cai06} who proposed to model 
\begin{align*}
& \mbox{TPR}^{\mathbb I}(c,s,t)=g_D\left(\eta_{\alpha}(t,s) + h(c)\right), \quad t\leq \tau \\
& \mbox{FPR}^{\mathbb S}(c,s,\tau)=g_{\tau}\left(\xi_{a}(s) + d(c)\right)
\end{align*}
where $g_D$ and $g_{\tau}$ are specified inverse link functions and $h(\cdot)$ and $d(\cdot)$ are unspecified baseline functions of threshold $c$. The dependence in time is parametrically modeled by $\eta_{\alpha}(t,s)=\alpha^{T}\eta(t,s)$ and $\xi_{a}(s)=a^{T}\xi(s)$ where $\eta(t,s)$ and $\xi(s)$ are vectors of polynomial or spline basis functions. These models are semiparametric with respect to the marker distribution in cases and nonparametric in regards to controls. As pointed out in \cite{Pepe08}, this model is very flexible as it does not specify any distributional form for the distribution of the marker given the event-time, but only model the effect of time-to-event on the marker distribution with a parametric form. Model estimation is performed by solving some estimating equations and large sample theory was established allowing a resampling method to construct confidence bands and make inference \cite{cai06}. Interestingly, the authors of \cite{cai06} also showed that covariates can easily be included in $\mbox{TPR}^{\mathbb I}$ and $\mbox{FPR}^{\mathbb S}$, enabling to directly quantify how performances of the marker vary with these covariates.

%
%
\section{Time-dependent AUC and Competing risks}\label{sec:AL}
We now consider the setting where a subject might experience multiple type of failures:  
in this section, we review extensions of time-dependent AUCs to competing risks.
For example, we may want to assess the discrimination of a  given score  on death from prostate cancer with death from other causes acting as a competing event.

\noindent For the sake of simplicity, we will assume there are only two competing events, and we let $\delta_{i}=j$ denote that subject $i$ experienced the  competing event of type $j$ ($j=1,2$, with $j=1$ for the event of interest). The observed data consists of a failure time and a failure type $(Z_i,\delta_i)$  with $\delta_i=0$ denoting a censored observation.

\noindent In their review paper  \cite{Pepe08} sketched most potential extensions and introduced event-specific sensitivity and specificity.
%
%
They also highlighted that the crucial point was to determine whether patients experimenting a competing event should be treated as  a control when evaluating the discrimination of the marker under study with respect to the event of interest.
More precisely, two settings can be considered. \vskip5pt

\noindent First, if marker $X$ is potentially discriminatory for both the event of interest and the competing event, then 
 both event specific AUCs should be considered simultaneously \cite{SahaHeagerty10,Foucher10}. 
 For illustration, in  the cumulative/dynamic setting, cases at time $t$ can be stratified according to the event type, $\mbox{Case}_1=\{i: T_i \leq t, \delta_i=1\}$ and   $\mbox{Case}_2 =  \{i: T_i \leq t, \delta_i=2\}$,  and controls at time $t$ are the event-free group at time $t$, Control=$\{i: T_i > t\}$.
 Following these lines Saha and Heargerty \cite{SahaHeagerty10}  proposed  event specific  versions of (\ref{cumTPRdynFPR})  
\begin{equation}
\mbox{TPR}_{j}^{\mathbb C}(c,t)=\P(X>c|T\leq t, \delta=j),\,\,\,
\mbox{FPR}^{\mathbb D}(c,t)=\P(X>c|T> t, \delta\in\{1,2\} ).
\label{cumTPRcrdynFPRcr}
\end{equation}
Estimation follows from \cite{heagerty00},  using the conditional cumulative incidence associated to competing event $j$, $\P(T \leq t, \delta=j | X)$, instead of the conditional survival function  of $\P(T\leq t | X)$.  In the context of renal transplantation, Foucher \emph{et al.} \cite{Foucher10} considered a slight modification of definitions (\ref{cumTPRcrdynFPRcr}), where  controls can also be ``event specific''. 
In addition, an extension of the incident/dynamic AUC to the competing events setting was proposed by Saha and Heargerty \cite{SahaHeagerty10}. \vskip5pt

\noindent The other option is to consider both event-free patients and patients with the competing event \cite{ZhengCai11,lee12} as controls. For instance, dynamic controls at time $t$ can be defined as the group $\{i: T_i>t\} \cup \{i: T_i\leq i, \delta_i = 2\}$. This leads to the estimation of only one ROC curve, for the event of interest.  In \cite{ZhengCai11}, Zheng \emph{et al.} based their approach on initial estimates of the conditional cumulative incidence function for the event of interest. Their initial method provides consistent estimators if  the proportional hazard assumption holds for each cause specific hazard. To relax this  assumption a  smooth estimator was also proposed.  Another approach was described in \cite{lee12}, which follows the lines of DeLong \emph{et al.} \cite{delong88}. 
However, the suitability of this method to deal with censored data is 
 not established.

\noindent We shall add that, as pointed out in  \cite{SahaHeagerty10,ZhengCai11}, employing a direct regression model for  the  conditional cumulative incidence  would lead to a simpler 
estimation of the cumulative/dynamic AUC and a less convoluted interpretation of the marker effect. However, the extension to the setting of a longitudinal marker \cite{cortese10} as well as the evaluation of a risk score (which is usually built with a cause-specific hazard approach) would not be straightforward. 

\section{Discussion}\label{sec:Discussion}

While the AUC is uniquely defined in the context of the evaluation of diagnostic tests, its extension to prognostic markers has led to the development of a variety of definitions:  
these definitions vary according to the underlying definitions of cases (incident or cumulative) and controls (static or dynamic),  and also depend on the study characteristics (the marker can be measured only once or repeatedly and competing events may be considered, or not). Regarding the choice of the retained definition for cases and controls, no clear guidance has really emerged in the literature. It seems however that the cumulative/dynamic definition may be more appropriate for clinical decisions making (enrollment in clinical trials for instance) while the incident/static definition may be more appropriate for ``pure'' evaluation of the marker (if interpretation of trends of AUC values over time is of particular interest for instance). Once this definition has been chosen, appropriate estimators are available, depending on various assumptions (independence of the marker and censoring, proportional hazards, ...), and we presented most of them in this review article.

\noindent For the sake of brevity, we were not able to cover some interesting extensions of time-dependent AUCs. In particular, covariate specific time-dependent ROC curves and AUCs have been studied in order to adjust the discrimination of a marker for external covariates (age, gender, ...). We refer the reader to \cite{CJS:CJS10046,Song08} for the standard setting, \cite{cai06} for the longitudinal setting and \cite{ZhengCai11} for the competing events setting. In addition, some authors advocate that not the entire ROC curve is of interest and the area under only a portion of it should be computed, leading to the so-called \emph{partial AUC} \cite{dodd2003partial}. In the context of prognostic markers, Hung and Chiang \cite{Hung11} proposed a nonparametric estimator of the cumulative/dynamic time-dependent version of the partial AUC. Other interesting extensions include diverse censoring patterns \cite{LiMa11} (only right-censoring was considered in this review) and the combination of results from multiple studies \cite{cai11} which is particularly useful in genomic studies.

\noindent Another closely related topic is the evaluation of the added predictive ability of a new marker: for instance, we may wonder how better a risk score would be if we added some biological markers (SNPs, genes, ...). We refer the reader to the works in \cite{chambless2011, demler2011equivalence, pencina2008evaluating, pepe2011testing} for some insights, noticing though that most of these works do not cover the right-censored setting considered in our review.

\bibliographystyle{plain}
\bibliography{survey2}
\end{document}